\documentclass[conference]{IEEEtran}
\IEEEoverridecommandlockouts

\usepackage{cite}
\usepackage{amsmath,amssymb,amsfonts}
\usepackage{algorithmic}
\usepackage{graphicx}
\usepackage{textcomp}
\usepackage{xcolor}
\usepackage{amsthm}
\newtheorem{definition}{Definition}
\usepackage{multicol, multirow, booktabs}

\def\BibTeX{{\rm B\kern-.05em{\sc i\kern-.025em b}\kern-.08em
    T\kern-.1667em\lower.7ex\hbox{E}\kern-.125emX}}

\usepackage{amsmath,amssymb,amsfonts,amsthm,optidef}
\usepackage{cite}
\usepackage{hyperref}
\usepackage{bm}
\usepackage{bbm}
\usepackage{xparse}
\usepackage{import}
\usepackage{algorithm}
\usepackage{algorithmic}
\usepackage[normalem]{ulem}

\makeatletter
\@ifpackageloaded{unicode-math}{%
	\newcommand{\mymathbold}{\symbf}%
}{%
	\usepackage{bm}%
	\newcommand{\mymathbold}{\bm}%
}
\makeatother

\newcommand{\scrbar}[1]{\overline{\mathcal{#1}}}

\ExplSyntaxOn

\cs_new_protected:Nn \bamboo_define:nnnnnN
{
	\cs_new_protected:cpx { #3 #1 #4 } { \exp_not:N #5{#6{#2}} }
}

\int_step_inline:nnn { `A } { `Z }
{
	\bamboo_define:nnnnnN
	{ \char_generate:nn { #1 } { 11 } }
	{ \char_generate:nn { #1 } { 11 } }
	{ bl }
	{ }
	{ \mymathbold }
	\use:n
	
	\bamboo_define:nnnnnN
	{ \char_generate:nn { #1 } { 11 } }
	{ \char_generate:nn { #1 } { 11 } }
	{ scr }
	{ }
	{ \mathcal }
	\use:n
	
	\bamboo_define:nnnnnN
	{ \char_generate:nn { #1 } { 11 } }
	{ \char_generate:nn { #1 } { 11 } }
	{ }
	{ hat }
	{ \widehat }
	\use:n
	
	\bamboo_define:nnnnnN
	{ \char_generate:nn { #1 } { 11 } }
	{ \char_generate:nn { #1 } { 11 } }
	{ }
	{ br }
	{ \overline }
	\use:n
	
	\bamboo_define:nnnnnN
	{ \char_generate:nn { #1 } { 11 } }
	{ \char_generate:nn { #1 } { 11 } }
	{ }
	{ tl }
	{ \widetilde }
	\use:n
	
	\bamboo_define:nnnnnN
	{ \char_generate:nn { #1 } { 11 } }
	{ \char_generate:nn { #1 } { 11 } }
	{ scr }
	{ br }
	{ \scrbar }
	\use:n
}
\int_step_inline:nnn { `a } { `z }
{
	\bamboo_define:nnnnnN
	{ \char_generate:nn { #1 } { 11 } }
	{ \char_generate:nn { #1 } { 11 } }
	{ bl }
	{ }
	{ \mymathbold }
	\use:n
	
	\bamboo_define:nnnnnN
	{ \char_generate:nn { #1 } { 11 } }
	{ \char_generate:nn { #1 } { 11 } }
	{ }
	{ hat }
	{ \hat }
	\use:n
	
	\bamboo_define:nnnnnN
	{ \char_generate:nn { #1 } { 11 } }
	{ \char_generate:nn { #1 } { 11 } }
	{ }
	{ br }
	{ \bar }
	\use:n
	
	\bamboo_define:nnnnnN
	{ \char_generate:nn { #1 } { 11 } }
	{ \char_generate:nn { #1 } { 11 } }
	{ }
	{ tl }
	{ \tilde }
	\use:n                            
}
\clist_map_inline:nn
{
	Gamma,Delta,Theta,Lambda,Xi,Pi,Sigma,Phi,Psi,Omega
}
{
	\bamboo_define:nnnnnN
	{ #1 }
	{ #1 }
	{ bl }
	{ }
	{ \mymathbold }
	\use:c
	
	\bamboo_define:nnnnnN
	{ #1 }
	{ #1 }
	{ op }
	{ }
	{ \op }
	\use:c
	
	\bamboo_define:nnnnnN
	{ #1 }
	{ #1 }
	{ scr }
	{ }
	{ \mathcal }
	\use:c
	
	\bamboo_define:nnnnnN
	{ #1 }
	{ #1 }
	{ }
	{ hat }
	{ \widehat }
	\use:c
	
	\bamboo_define:nnnnnN
	{ #1 }
	{ #1 }
	{ }
	{ br }
	{ \overline }
	\use:c
	
	\bamboo_define:nnnnnN
	{ #1 }
	{ #1 }
	{ }
	{ tl }
	{ \widetilde }
	\use:c
}
\clist_map_inline:nn
{
	alpha,beta,gamma,delta,epsilon,zeta,eta,theta,iota,kappa,
	lambda,mu,nu,xi,pi,rho,sigma,tau,phi,chi,psi,omega
}
{
	\bamboo_define:nnnnnN
	{ #1 }
	{ #1 }
	{ bl }
	{ }
	{ \mymathbold }
	\use:c
	
	\bamboo_define:nnnnnN
	{ #1 }
	{ #1 }
	{ }
	{ hat }
	{ \hat }
	\use:c
	
	\bamboo_define:nnnnnN
	{ #1 }
	{ #1 }
	{ }
	{ br }
	{ \bar }
	\use:c
	
	\bamboo_define:nnnnnN
	{ #1 }
	{ #1 }
	{ }
	{ tl }
	{ \tilde }
	\use:c
}

\ExplSyntaxOff

\DeclarePairedDelimiter{\norm}{\lVert}{\rVert}

\DeclarePairedDelimiterX{\ip}[2]{\langle}{\rangle}{#1,#2}

\DeclarePairedDelimiterXPP{\normsub}[2]{}{\lVert}{\rVert}{_{#2}}{#1}
\DeclarePairedDelimiterXPP{\ipsub}[3]{}{\langle}{\rangle}{_{#3}}{#1,#2}

\DeclarePairedDelimiterXPP{\ipHS}[2]{}{\langle}{\rangle}{_{\mathrm{HS}}}{#1, #2}
\DeclarePairedDelimiterXPP{\normHS}[1]{}{\lVert}{\rVert}{_{\mathrm{HS}}}{#1}

\DeclarePairedDelimiterXPP{\ipF}[2]{}{\langle}{\rangle}{_{\mathrm{F}}}{#1, #2}
\DeclarePairedDelimiterXPP{\normF}[1]{}{\lVert}{\rVert}{_{\mathrm{F}}}{#1}

\DeclarePairedDelimiterXPP{\normt}[1]{}{\lVert}{\rVert}{_{2}}{#1}
\DeclarePairedDelimiterXPP{\normo}[1]{}{\lVert}{\rVert}{_{\mathrm{1}}}{#1}

\DeclarePairedDelimiterXPP{\dkl}[2]{\operatorname{D_{KL}}}{(}{)}{}{#1 \: \delimsize\Vert \: #2}

\DeclarePairedDelimiterXPP{\restr}[2]{}{{}}{\vert}{_{#2}}{#1}

\newcommand{\R}{\mathbb{R}}

\begin{document}
\title{MANGO: Learning Disentangled Image Transformation Manifolds with Grouped Operators
}

\author{
\IEEEauthorblockN{
Brighton Ancelin\textsuperscript{1}, Yenho Chen\textsuperscript{1}, Alex Saad-Falcon\textsuperscript{1}, Peimeng Guan\textsuperscript{2}, Chiraag Kaushik\textsuperscript{2}, \\
Nakul Singh\textsuperscript{1}, and Belen Martin-Urcelay\textsuperscript{2}} 
\IEEEauthorblockA{
\textsuperscript{1}\textit{ML@GT} and
\textsuperscript{2}\textit{Dept. of Electrical and Computer Engineering}\\
\textit{Georgia Institute of Technology, Atlanta, GA}\\
\{ \tt bancelin3, yenho, asf3, pguan6, ckaushik7, nsingh360, burcelay3\}@gatech.edu}
\thanks{This work was supported in part by the James S. McDonnell Foundation grant
number 220020399, the Georgia Institute of Technology, and the UR2PhD program of the Computing Research Association. The student authors are listed in alphabetical order by last name.}
}

\maketitle
\begin{abstract}
Learning semantically meaningful image transformations (i.e. rotation, thickness, blur) directly from examples can be a challenging task. Recently, the Manifold Autoencoder (MAE) \cite{connor2020representing} proposed using a set of Lie group operators to learn image transformations directly from examples. However, this approach has limitations, as the learned operators are not guaranteed to be disentangled and the training routine is prohibitively expensive when scaling up the model. To address these limitations, we propose MANGO (transformation Manifolds with Grouped Operators) for learning disentangled operators that describe image transformations in distinct latent subspaces. Moreover, our approach allows practitioners the ability to define which transformations they aim to model, thus improving the semantic meaning of the learned operators. Through our experiments, we demonstrate that MANGO enables composition of image transformations and introduces a one-phase training routine that leads to a $100\times$ speedup over prior works.
\end{abstract}
\begin{IEEEkeywords}
Image transformations, disentangled representation, autoencoder, generative model.
\end{IEEEkeywords}
\section{Introduction}
In many scientific domains, learning semantically meaningful transformations in high-dimensional image datasets can improve our understanding of complex patterns within the underlying system. For example, in medicine, learning image transformations between MRI scans of sick and healthy patients can provide insights into disease mechanisms. To ensure that these learned transformations are interpretable, the resulting representations must be low-dimensional and disentangled.

Although there exists numerous methods to learn a low-dimensional representation of high-dimensional image manifolds\cite{Wang_2014_CVPR_Workshops, 8417906, GAN, ni2022manifoldlearningbenefitsgans, kingma2022autoencodingvariationalbayes, rezende2014stochasticbackpropagationapproximateinference}, few works aim to model semantically meaningful transformations between different points on this manifold. Many latent variable models incorrectly assume Euclidean transformations in the latent space, which can be inappropriate when modeling certain transformations that require forming a closed path, such as image rotations.

Recently, a new line of research \cite{ni2022manifoldlearningbenefitsgans, fallah2023manifold, kim2018disentangling} has imposed structure by constraining image transformation onto a low-dimensional manifold that can be learned. In particular, the Manifold Autoencoder (MAE) \cite{connor2020representing} learns a dictionary of Lie group operators, referred to as \emph{transport operators}, which can be linearly combined to model possible transformations between latent points. To encourage the learning of semantically meaningful and disentangled operators, the MAE optimizes an objective function that promotes sparsity of the weights and shrinkage of the operators.

Despite demonstrating improved extrapolation along transformation paths, there are two critical limitations of the MAE that prevent it from efficiently learning a disentangled latent space. First, the MAE objective does not guarantee that latent transformations are orthogonal which often results in the learning of overlapping transport operators for distinct transformations.   
Second, training MAE requires an expensive three-phase procedure which becomes prohibitively expensive when scaling up the model.  

To address these limitations of MAE, we propose  {\bf MANGO}, an approach for learning disentangled image transformation {\bf Man}ifolds using {\bf G}rouped {\bf O}perators. We introduce the concept of disentangled operators and enforce disentanglement by constraining the action of each operator to a distinct latent subspace. Furthermore, our method allows practitioners to specify which semantically meaningful transformations they aim to model. Additionally, we show that our disentangled formulation enables composable latent transformations that can generate realistic images, even for out-of-distribution transformations. Finally, we present a one-phase training strategy that significantly improves computational efficiency, demonstrating a $100\times$ speedup over previous methods. 

\begin{figure}[htp]
    \centering
    \includegraphics[width=1\linewidth]{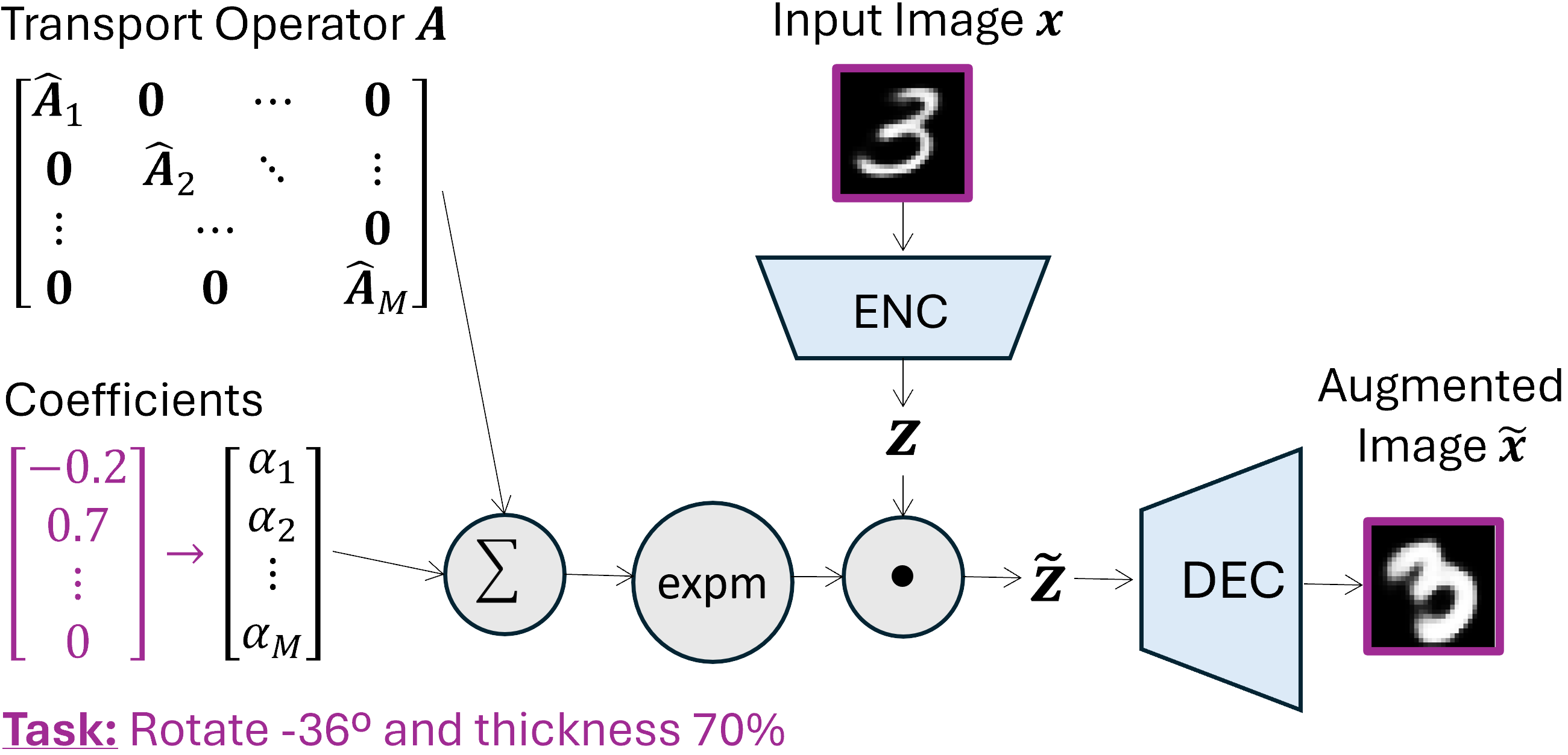}
    \caption{Block diagram of MANGO. The model simultaneously learns a transport operator $\blA$, where each block diagonal component represents a semantically meaningful transformation, and an autoencoder that reconstructs the transformed image.}
    \label{fig:block_diagram}
\end{figure}
\section{Background and Related Work}

\subsection{Disentangled Representation Learning}
Disentangled representation learning seeks to identify independent latent factors of variation that best describe the dataset of interest. By doing so, disentangled models improve the interpretability of the latent variables which encourages the discovery of semantically meaningful structures. Various techniques in deep learning literature have been developed to promote the learning of disentangled latent spaces. For example, $\beta$-VAE \cite{burgess2018understanding} applies strong regularization to the KL divergence term in the evidence lower bound to encourage the latent factor distribution to align with an isotropic Gaussian prior. Similarly, FactorVAE \cite{kim2018disentangling} promotes disentanglement by directly penalizing the total correlation of the latent prior. Alternatively, Generative Adversarial Networks (GANs), which aim to learn generative models capable of producing samples indistinguishable from real data by a discriminative classifier, provide a different strategy to disentanglement. In particular, InfoGAN \cite{chen2016infogan} achieves disentanglement by maximizing the mutual information between structured latent subspaces and the generation process. Other line of work learns latent representations in an unsupervised manner \cite{zimmermann2021contrastive, cheng2020club, oord2018representation}, where \cite{cheng2020club} minimizes the mutual information between the content embedding and domain embedding to encourage independence. 

In general, however, these methods do not guarantee that the learned latent factors will correspond to semantically meaningful transformations, since they lack mechanisms to incorporate additional practitioner-provided information. Furthermore, these models assume a Euclidean structure in the latent space, which may be unsuitable for certain image transformations that follow closed paths, such as rotations where transformations are more appropriately modeled on $SO(\cdot)$ manifolds.

\subsection{Transport Operators}
Transport operators \cite{culpepper2009learning} offer a framework for modeling continuous transformations between high-dimensional data points in their original space, $\blx \in \mathbb{R}^D$, by defining transformations to follow the flow of a linear dynamical systems, $\dot{\blx} = \blA \blx$. Given an initial point $\bm{x}_0$ and a linear operator $\blA \in \mathbb{R}^{D \times D}$, the trajectory of the transformation is given by $\bm{x}_t = {\rm expm} (t \blA) \bm{x}_0$ for all time $t\in \mathbb{R}$, where ${\rm expm}$ is the matrix exponential. Many works \cite{mudrik2024decomposed, chau2020disentangling, sohl2010unsupervised, chen2024probabilistic, mudrik2024linocs} further decompose $\blA$ into a linear combination of $M$ {\emph transport operators} $\{\bm{A}_m \}_{m=1}^M$ such that, $\blA = \sum_{m=1}^M c_m \blA_m$,
where $c_m \in \mathbb{R}$ is a coefficient that determines the contribution of each operator in a particular transformation. The set of all possible transport operators define a Lie group \cite{boothby2003introduction} and can efficiently represent the manifolds surface. 

To encourage the learning of statistically independent transport operators, \cite{culpepper2009learning} proposes to promote sparsity in the coefficients and shrinkage over the transport operators through regularized optimization of $\bm{A}, \bm{c}$ over the following objective,
\begin{equation}
    \label{eq:transop}
    \frac12 \normt*{\tilde{\blx} - \text{expm}\left(\sum_{m=1}^M c_m \blA_m \right) \blx}^2 + \frac\gamma2 \sum_{m=1}^M \norm{\blA_m}_\mathrm{F}^2 + \zeta \norm{\bm{c}}_1
\end{equation}
where $\gamma$ and $\zeta$ are penalty weights for the operator shrinkage and the amount of coefficient sparsity respectively. Since concurrently learning both transport operators and the coefficients can lead to training instability, prior works instead alternate between updating $\bm{A}$ and $\bm{c}$.

\subsection{Manifold Autoencoder (MAE)}
Learning transport operators in the high-dimensional data space can be challenging numerically and computationally expensive as a result of the matrix exponential. To address this limitation, the manifold autoencoder (MAE) \cite{connor2020representing, connor2021variational} proposes to learn transport operators in a low-dimension latent space of an autoencoder. This is accomplished through a three-phase training routine. First, an autoencoder is trained with the basic reconstruction loss. Second, the algorithm fixes the autoencoder weights and trains the transport operators (in the latent space) using pairs of neighboring points with loss function (\ref{eq:transop}) (modified so that data $\blx,\tilde{\blx} \in \R^D$ are replaced by latent representations $\blz,\tilde{\blz} \in \R^L$). 
Third, the algorithm prunes irrelevant operators and simultaneously fine-tunes the autoencoder weights alongside the remaining transport operators.  Although MAE improves the computational efficiency of transport operators by learning transformations in the latent space, it still relies on an expensive inner optimization procedure as a result of the $\ell_1$ term in Equation (\ref{eq:transop}). Additionally, the learned latent transformations are not guaranteed to be disentangled since the inferred active support set between different transformation may use overlapping coefficients.

\section{Methodology}
This Section describes the main components of MANGO. Figure \ref{fig:block_diagram} illustrates our approach.

\subsection{Learning a Low-Dimensional Latent Manifold}
For a given dataset $\{\blx_i\}_{i=1}^N$, let $\scrH \coloneqq \{h_{m, \alpha} \colon \R^D \to \R^D\}_{m=1}^M$ be a set of continuous semantic transformations on the data, where each $h_m$ is parameterized by a scalar $\alpha \in [-1, 1]$. We train an autoencoder (with encoder and decoder denoted $f$ and $g$, respectively) to learn a low-dimensional latent representation $\blz_i \in \R^L$ of the $\blx_i \in \R^D$ such that the transformations of $\scrH$ are represented in the latent space (approximately) by a structured class of manifolds.

\subsection{Enforcing Disentangled Operators with Group Structure}
Building on prior works \cite{bengio2013representation, wang2022disentangled}, we propose the following definition of a disentangled representation for operators,
\begin{definition} (Disentangled Operators)
  A set of operators $\{ \bm{A}_m\}_{m=1}^M$ are disentangled if every pair of operators $\bm{A}_i$ and $\bm{A}_j$, where $i \neq j$, satisfy $\langle \bm{A}_i, \bm{A}_j \rangle := \operatorname{trace}(\bm{A}_j^{\dagger}\bm{A}_i)= 0$. 
\end{definition}
For each transformation (indexed $m  = 1, \dots, M$), we learn the transport operator $\blA_m$ by attempting to minimize
\[
    T_m = \normt{\tilde{\blz} - \text{expm}(\alpha \blA_m) \blz}^2 + \gamma \norm{\blA_m}_F^2
\]
where $\blz \coloneqq f(\blx)$ and $\tilde{\blz} \coloneqq f(\tilde{\blx})$ are the latent representations corresponding to the original sample $\blx$ and the transformed sample $\tilde{\blx} \coloneqq h_{m, \alpha}(\blx)$, respectively. This formulation encourages the operator to define a continuous transformation path in the latent space parameterized by $\alpha$. We further constrain $\blA$ to be a block-diagonal matrix, where each $\blA_m$ occupies a distinct block with no overlapping support, ensuring \textit{disentanglement}. This has the effect that the action of the $m^{\text{th}}$ transport map on $\blz$ is localized to a small, identifiable subset of coordinates.

\subsection{Improving Training Efficiency with One-Phase Approach}

Combining these two components, we obtain an overall loss value for each pair of transformed points in the input space $(\blx, \tilde{\blx})$ and their corresponding latent representations $(\blz, \tilde{\blz})$:
\begin{equation}\label{eq:loss}
E = \normt{\blx - g(f(\blx))}^2 + \normt{\tilde{\blx} - g(f(\tilde{\blx}))}^2 + \lambda T_m,
\end{equation}
where the the first two terms encourage accurate reconstruction of the original and transformed inputs, while the final term ensures that the block-diagonal transport operator $\blA_m$ transforms one latent representation into the corresponding transformed latent representation. The training methodology is summarized in Algorithm \ref{alg:alg}.
\begin{algorithm}[t]
  \caption{MANGO Algorithm}\label{alg:alg}
  \begin{algorithmic}[1]
    \STATE \textbf{Input: }Randomly initialized operators $\{\blA_m\}$ and network weights $\bltheta$
    \STATE \textbf{Output: } Learned transport operators $\{\blA_m\}$ and autoencoder weights $\bltheta$
    \FOR{$t = 1, 2, \dots$}
        \STATE Sample data batch $\scrB = \{\blx_1, \dots, \blx_B\}$
        \STATE Encode batch to obtain $\scrL = \{\blz_1, \dots, \blz_B\}$
        \FOR{$m = 1$ \TO $M$}
            \STATE $\alpha \sim \text{Unif}([-1,1])$
            \STATE Apply $h_{m, \alpha}$ to $\scrB$ to obtain $\tilde{\scrB}_m = \{\tilde{\blx}_1, \dots, \tilde{\blx}_B\}$
            \STATE Encode $\tilde{\scrB}_m$ to obtain $\tilde{\scrL}_m = \{\tilde{\blz}_1, \dots, \tilde{\blz}_B \}$
            \STATE $\blA_m \gets \blA_m - \eta_1 \sum_{i=1}^B \frac{\partial E_i}{\partial \blA_m}$
        \ENDFOR
        \STATE $\bltheta \gets \bltheta - \eta_2 \sum_{i=1}^B \frac{\partial E_i}{\partial \bltheta}$
    \ENDFOR
  \end{algorithmic}
\end{algorithm}

\section{Results}
The MNIST handwritten digits dataset \cite{lecun1998gradient} is a widely used benchmark for evaluating latent structures \cite{connor2020representing, connor2021variational} because it allows for straightforward assessment of semantically meaningful transformations through standard image operations (e.g., rotations, thickness, blurriness). We use MNIST to assess MANGO's ability to learn disentangled latent operators. As a result of enforcing disentanglement, we demonstrate that we can linearly combine the learned operators to perform multiple semantic effects simultaneously. Additionally, we compare the computational complexity of MANGO's one-phase training procedure with the MAE's three-phase training procedure, demonstrating a substantial improvement in training runtime.
\begin{figure}[t]
    \centering
    \includegraphics[width=0.9\linewidth]{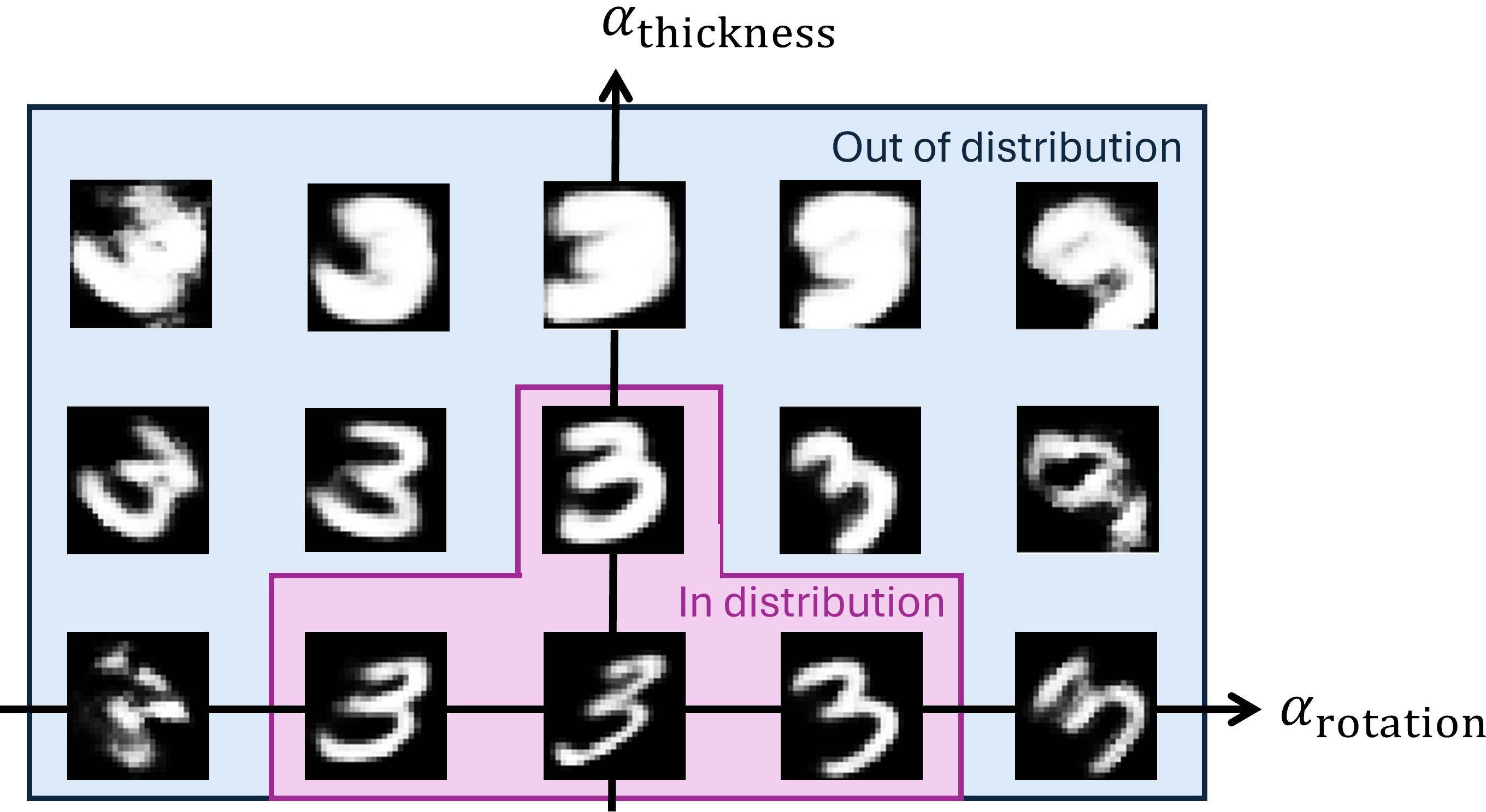}
    \caption{Augmentations with various combinations of rotations and thickness changes.}
    \label{fig:results_augmentations}
\end{figure}

\begin{figure*}[t!]
    \centering
    \begin{minipage}{0.6\linewidth} 
        \centering
        \includegraphics[width=\linewidth]{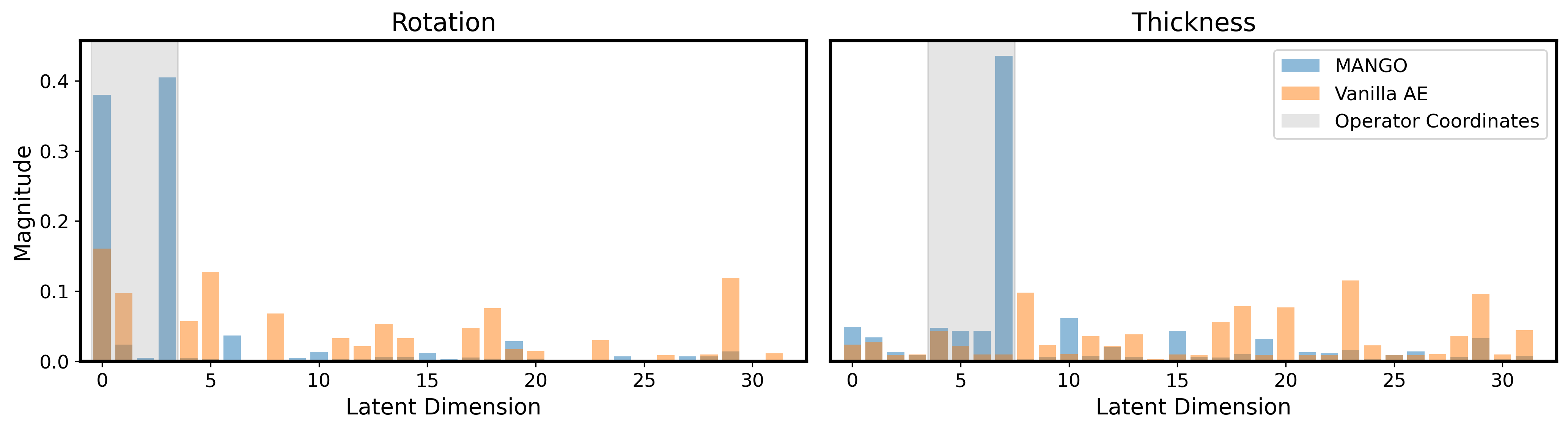}
        \vspace{-20pt}
        \caption{MANGO achieves a neatly disentangled latent space. The figure shows the magnitude of each coordinate in the first principal component for both models. MANGO exhibits stronger concentration and alignment with learned operator coordinates.}
        \label{fig:pca}
    \end{minipage}
    \hfill
    \begin{minipage}{0.38\linewidth} 
        \centering
        \vspace{8pt}
        \scriptsize 
        \begin{tabular}{|c|c|c|c|c|}
            \hline
            Latent dimension $L$ & 16 & 32 & 64 & 128 \\
            \hline
            MAE & 6.90 & 20.52 & 72.54 & 319.50 \\
            MANGO & 0.18 & 0.18 & 0.20 & 0.20 \\
            \hline
        \end{tabular}
        
        \vspace{8pt} 
        
        \begin{tabular}{|c|c|c|c|c|}
            \hline
            Dictionary Size $M$ & 2 & 4 & 6 & 8 \\
            \hline
            MAE & 8.98 & 12.17 & 14.50 & 19.92 \\
            MANGO & 0.16 & 0.16 & 0.18 & 0.18 \\
            \hline
        \end{tabular}
        
        \vspace{14pt}    
        \caption{Training runtimes (in seconds) per batch (of size 64) for fixed dictionary size $M=8$ and for fixed latent dimension $L=32$.}
        \label{tab:time_vs_latent_size_and_dictionary}
    \end{minipage}
\end{figure*}

\subsection{Disentangled Operators are Composable}
We empirically show that MANGO leverages the disentangled manifold structure to learn semantically meaningful operators on the MNIST dataset. Figure \ref{fig:results_augmentations} shows augmented images generated by applying two distinct transport operators. One operator corresponds to a rotation transformation, while the other increases the thickness of the digits. Note that the reconstruction performance remains comparable to the baseline autoencoders. MANGO provides the additional benefit of interpretability, without a substantial trade-off in image quality.

\begin{figure}
    \centering
    \includegraphics[width=0.75\linewidth]{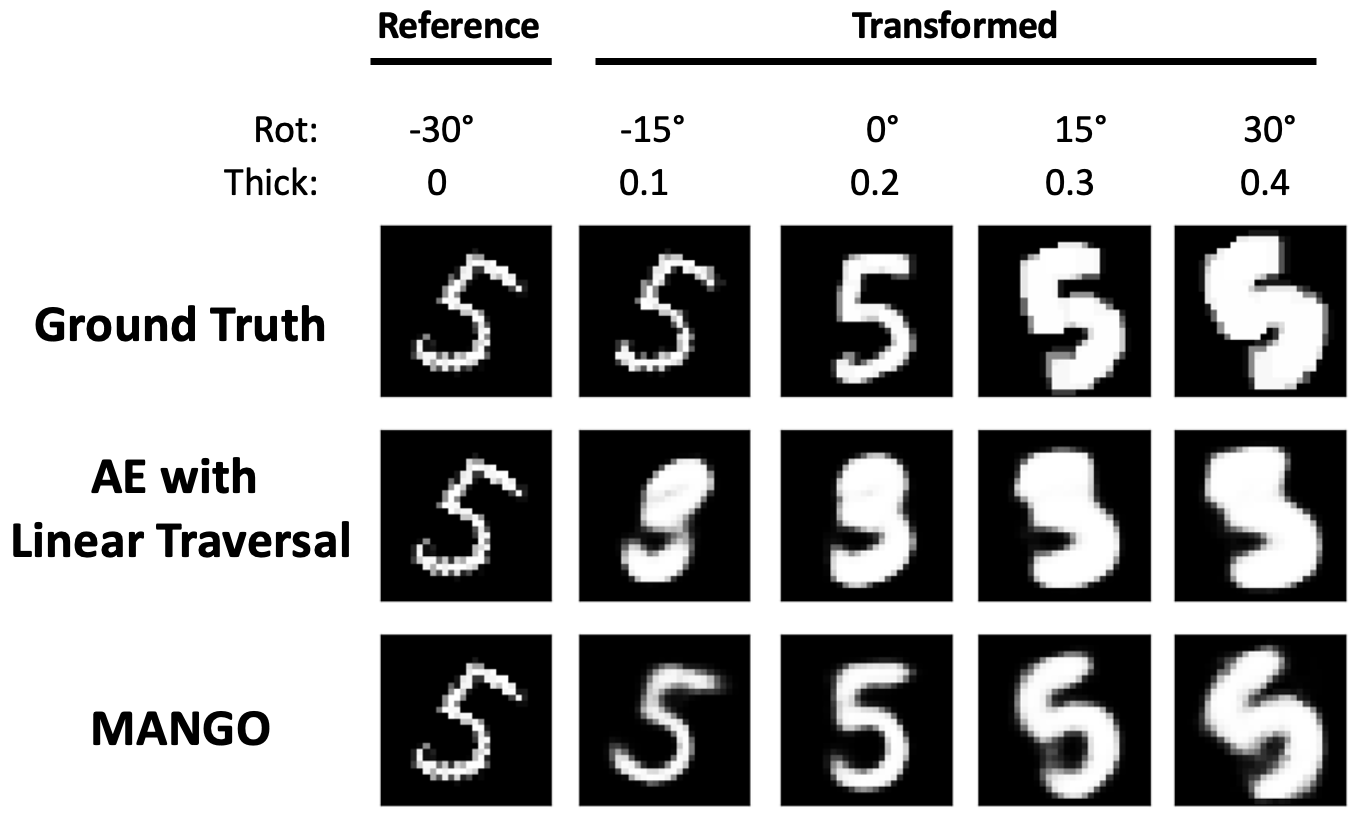}
    \caption{Comparison of image transformations. MANGO transformations retain image identity unlike the AE.}
    \label{fig:MNISTfive}
\end{figure}

Additionally,  the operators learned by MANGO generalize beyond the training dataset in two ways. First, they are able to transport images further than the transformations observed during training, showing robustness in extrapolation. Second, MANGO is able to linearly combine the learned operators to achieve complex transformations. For instance, as illustrated in Figures \ref{fig:results_augmentations} and \ref{fig:MNISTfive}, the model successfully generates augmented images where the digits are both rotated and thickened simultaneously. This demonstrates the model’s ability to compose transformations in a meaningful and interpretable manner, in contrast with vanilla AEs where images lose their identity.

Figure \ref{table:recon_loss} includes quantitative metrics for our MNIST experiments. We measure 1) image reconstruction error to evaluate the quality of the autoencoder and 2) transformed image reconstruction error to assess the quality of the latent transformation function. We report MSE and LPIPS \cite{zhang2018unreasonable}. For the image transformation metric, we obtain latent traversals from a vanilla autoencoder (AE) by fitting linear transformations on latent points and applying these transformations iteratively to a reference image's embedding. Although AE slightly outperforms MANGO in reconstructing available images, it suffers when generating image transformations. In fact, MANGO demonstrates a 60\% improvement in transformed reconstructions.

\begin{figure}[]
    \centering
    \setlength{\tabcolsep}{.7pt}
    \begin{tabular}{cc|cc|cc|c}
                 &      & \multicolumn{2}{c|}{Image} & \multicolumn{2}{c|}{Transformed} & \multicolumn{1}{c}{Disent.} \\ 
                         Models  &  $\alpha$  & MSE   & LPIPS & MSE  & LPIPS & MIG  \\ 
                           \hline
    \multirow{3}{*}{AE}   & rotate &    {\bf 0.011}      &    {\bf 0.043}   &     0.072     &    0.119   &   0.005 \\
                           & thick &    {\bf 0.007}      &     {\bf 0.020}  &     0.093     &    0.082    &  0.034\\ 
                  & rotate + thick &    {\bf 0.013}      &   {\bf 0.042}     &    0.076      &      0.085   &     -  \\ 
                           
                           \hline
    \multirow{3}{*}{MANGO}   & rotate &    0.015      &    0.051    &      {\bf 0.027}   &     {\bf 0.057}     &     {\bf 0.031}    \\
                           & thick &    0.011      &    0.030     &       {\bf 0.022}   &     {\bf 0.039}     &       {\bf 0.11}  \\ 
    
                          & rotate + thick &   0.018      &    0.053    &     {\bf 0.040}   &     {\bf 0.067}     &       \\ 
                           
    \end{tabular}
    \caption{Quantiative metrics for the MNIST experiments. We compute scores that quantify image reconstruction, image transformations, and disentanglement. Lower is better for MSE and LPIPS while higher is better for MIG.}
    \label{table:recon_loss}
\end{figure}

\subsection{Grouped Operators Improve Training Time}
The disentangled group structure of MANGO allows for simpler backpropagation computations, leading to a faster training process. During training for our MNIST experiments, we observe that MANGO takes 12 minutes to converge while the baseline MAE requires 138 hours to converge. The overall algorithm takes \boldmath $0.14\%$ of the time to fully converge.

The autoencoders in both approaches are fully connected neural networks with hidden layer sizes (256, 64, $L$ (latent space), 64, 256), leaky ReLU hidden layer activations, and a sigmoid final layer activation. The first table in Figure \ref{tab:time_vs_latent_size_and_dictionary} compares the runtimes of batch processing for different latent dimensions $L$; while MANGO requires similar computation time for all $L$, MAE quickly scales on order roughly $L^2$. This is due to the order $L$ (with small coefficient) scaling of the block diagonal MANGO transport matrices, whereas the dense transport matrices of MAE scale on order $L^2$. As a result MANGO trains up to $1500\times$ faster than MAE for a latent space of size 128. The second table in Figure \ref{tab:time_vs_latent_size_and_dictionary} compares the runtimes for different dictionary sizes $M$; again, MANGO requires similar computation time for all $M$ while MAE scales poorly. For reasonable $L$ and $M$, the manifold-relevant computations are nearly negligible compared to the neural network computations, and as such we observe the nearly constant runtimes for varying $L$ and $M$ on MANGO. 
All experiments were run on an AMD Ryzen 3900 12 core processor and NVIDIA GeForce RTX 2070.

\subsection{MANGO Disentangles the Latent Space}
To study MANGO's effect on the latent space, we apply principal component analysis (PCA) to latent representations of image augmentations. We select 10 random images from the dataset and generate 100 augmentations for each by varying rotation and thickness. These augmentations are then fed to both MANGO and the vanilla autoencoder for comparison. We apply PCA to each set of augmentations and average the results over the 10 sets. For both models, the explained variance concentrates in the first few singular components, with MANGO showing slightly better concentration. Crucially, the large coordinates of MANGO's leading eigenvector align with the operator coordinates, indicating a disentangled latent space. In contrast, the vanilla autoencoder's energy is spread across many coordinates. Figure \ref{fig:pca} illustrates these results. 

We also measure disentanglement quantitatively with the MIG score \cite{chen2018isolating}. the results in Figure \ref{table:recon_loss} show that MANGO achieves superior disentanglement compared to AE.

\section{Future Work}
\begin{figure}[H]
    \centering
    \includegraphics[width=0.9\linewidth]{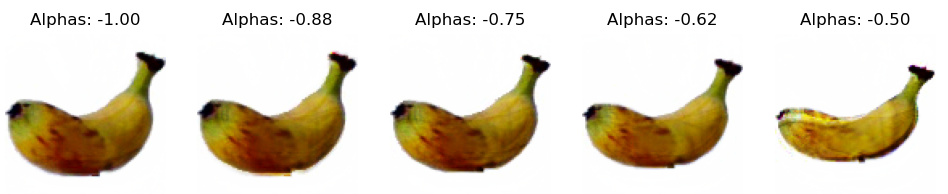}
    \caption{Generated images of rotated bananas, for rotation angles not seen in the training set.}
    \label{fig:banana}
  \end{figure}

MANGO provides a general framework which can in principle apply to complex image transformations. As future work, we plan on extending our experimental results to learning image transformation in the Fruits-360 dataset \cite{muresan2018fruit}, a public dataset of images of rotated fruit. This dataset is significantly more challenging than MNIST since our models must represent 3D transformations given 2D snapshots. We train on a small range of rotation parameters ($\alpha$), and find that the learned transport operator can in fact generate images of rotated bananas for values outside the training set. These preliminary results, shown in Figure \ref{fig:banana}, indicate that our framework could be used for more challenging transformation learning.
\vspace{-0.5em}

\bibliographystyle{unsrt}
\bibliography{refs.bib}
\end{document}